\documentclass[journal]{IEEEtran}

\usepackage[english]{babel}
\usepackage{mathptmx}

\usepackage[letterpaper,top=2cm,bottom=2cm,left=3cm,right=3cm,marginparwidth=1.75cm]{geometry}

\usepackage{amsmath}
\usepackage{graphicx}
\usepackage[colorlinks=true, allcolors=black]{hyperref}
\usepackage{subcaption}
\usepackage{float}
\usepackage[normalem]{ulem}
\usepackage{cite}
\usepackage{amssymb}

\begin{document}

\title{Perceptive, Resilient, and Efficient Networks assisted by Reconfigurable Intelligent Surfaces}

\author{\IEEEauthorblockN{Giorgos Stratidakis,
Sotiris Droulias, and \\
Angeliki Alexiou,~\IEEEmembership{Member,~IEEE}}
}

\maketitle
\begin{abstract}
Wireless communications are nowadays shifting to higher operation frequencies with the aim to meet the ever-increasing demand for bandwidth. While reconfigurable intelligent surfaces (RISs) are usually envisioned to restore the line-of-sight of blocked links and to efficiently counteract the increased pathloss, their functionalities can extend far beyond these basic operations. Owing to their large surface and the multitude of scatterers, RISs can be exploited to perform advanced wavefront engineering, essentially transforming the incident beam into a non-trivial reflected beam that is able to address the challenges of high frequencies more efficiently than conventional beam-forming. In this paper it is demonstrated how advanced wavefront engineering with RISs enables beam profiles that are able to focus, bend and self-heal, thus offering functionalities beyond the current state-of-the-art. Their potential as enablers of perceptive, resilient, and efficient networks is discussed, and a localization technique based on a hybrid beam-forming/beam-focusing scheme is demonstrated.

\end{abstract}
\begin{IEEEkeywords}
Reconfigurable intelligent surfaces, Advanced wavefront engineering, Beam-forming, Beam-focusing, Self-healing, Self-accelerating
\end{IEEEkeywords}
\section{Introduction}

\noindent In recent years, research in industry and academia has been directed towards identifying critical scenarios for future networks and devising the corresponding enablers to address the associated challenges. To take full advantage of the large bandwidth offered by high frequency bands, and to ensure high quality links, it is necessary to efficiently counteract the sources of blockage and attenuation that lead to signal degradation and, therefore, the necessity for new approaches to address such challenges becomes vital \cite{Pan2021}. In this context, the capability to flexibly engineer the wavefront of beams that carry the information, taking into consideration environmental parameters and usage scenario characteristics, opens up a new ‘beyond communications’ playground, where communication nodes can be employed to identify the shape of surrounding obstacles, map the environment, detect the presence of objects, focus power, localize, track mobility and navigate \cite{Khalily2022}. \\
\indent It is gradually becoming common ground that future networks will be programmable, reconfigurable, with environmental awareness that maps the communication landscape, achieving unprecedented quality of service. As a result, they are expected to support a multitude of functionalities and, therefore, to bear multiple qualities; they are expected to be  \textbf{perceptive} within the communication environment, \textbf{resilient} to communication outages and power \textbf{efficient}. Specifically, a perceptive network will be able to detect changes in the communication environment, such as the motion of the users or objects that block active links and even predict their motion. A resilient network will be able to restore quickly the connections in cases of outage from power loss or blockage and even avoid blockage. The link restoration will be seamless so as not to affect the user experience. A power efficient network will function with low energy requirements, and with as few corrective measures as possible. \\
\indent These qualities can be enabled with beam profiles that offer advanced functionalities beyond conventional beam-forming. For example, beams that are able to redistribute the incident power into a focused area, or even bend or self-heal, are ideal candidates. They can be produced either directly from a specially designed emitter, e.g. a large antenna array, or a reflecting surface that is able to transform the incoming beam, such as a reconfigurable intelligent surface (RIS). Such beams usually involve rapidly varying phase profiles and, therefore, to produce them it is required that their wavefront is adequately sampled. As a result, many radiating elements are usually necessary, rendering large antenna arrays that require bulky hardware non-practical and inefficient. \\
\indent Contrary to large antenna arrays, RISs are nearly passive surfaces, that is, they do not require active power amplifiers, digital signal processing, and multiple RF chains \cite{Linlong2020}. Owing to its passivity, the RIS may offer hundreds or thousands of elements at relatively low power consumption levels, making it ideal for advanced wavefront engineering operations that require dense wavefront sampling. Besides promising to restore the line-of-sight (LoS) of links interrupted by a blocker and to increase the channel gain via beam forming, the potentially large size of RISs enables a number applications, such as cost-efficient redirection of a beam, focusing the power of a beam to a small area with controllable size, generating beams that can bend to bypass obstacles, enabling them to reach users beyond the LoS of conventional beams. Overall, RISs can adjust the wireless medium, which was previously thought to be uncontrollable, by controlling the scattering, refraction and reflection characteristics of waves \cite{Basar2019}. With RISs, the wavefront engineering also becomes programmable \cite{Droulias2022}. This enables the transmitter to sense the wireless environment and allows for integrated communication and sensing, thus enabling full advantage of the high bandwidth offered by high frequency bands, and promising high quality, perceptive, resilient and efficient links. \\
\indent So far, communications have been driven by omnidirectional emitters and directional beamformers, while, in recent years, beam-focusing is gaining ground. Additionally, most applications in wireless communications have been designed for the far-field of transmitting antennas. However, with RISs it is now possible to adjust the near- and far-field regions \cite{Stratidakis2021a}, thus enabling new functionalities. In this work. In this paper, the potential of RISs for advanced wavefront engineering is investigated. The main contributions of this paper are as follows.
\begin{itemize}
    \item The concept of beam-forming, beam-focusing, self-accelerating and self-healing as enablers of perceptive, resilient, and efficient networks is proposed.

    \item It is demonstrated how beams that are able to focus, bend and self-heal enable advanced functionalities necessary for future networks beyond typical beam-forming, and the design specifications to reach their full potential are discussed.

    \item A hierarchical localization method based on a hybrid beam-forming/beam-focusing scheme is proposed.

    \item New approaches to counteract the effect of blockage based on self-accelerating and self-healing beams are proposed.

    \item Methods for achieving energy efficient networks are demonstrated. The concept of wireless power transfer and how the received power can be maximized with low transmitted power is discussed.
\end{itemize}
%
%
%
%
\section{Advanced wavefront engineering and capabilities}
%
\begin{figure}
\centering
\includegraphics[width=\linewidth]{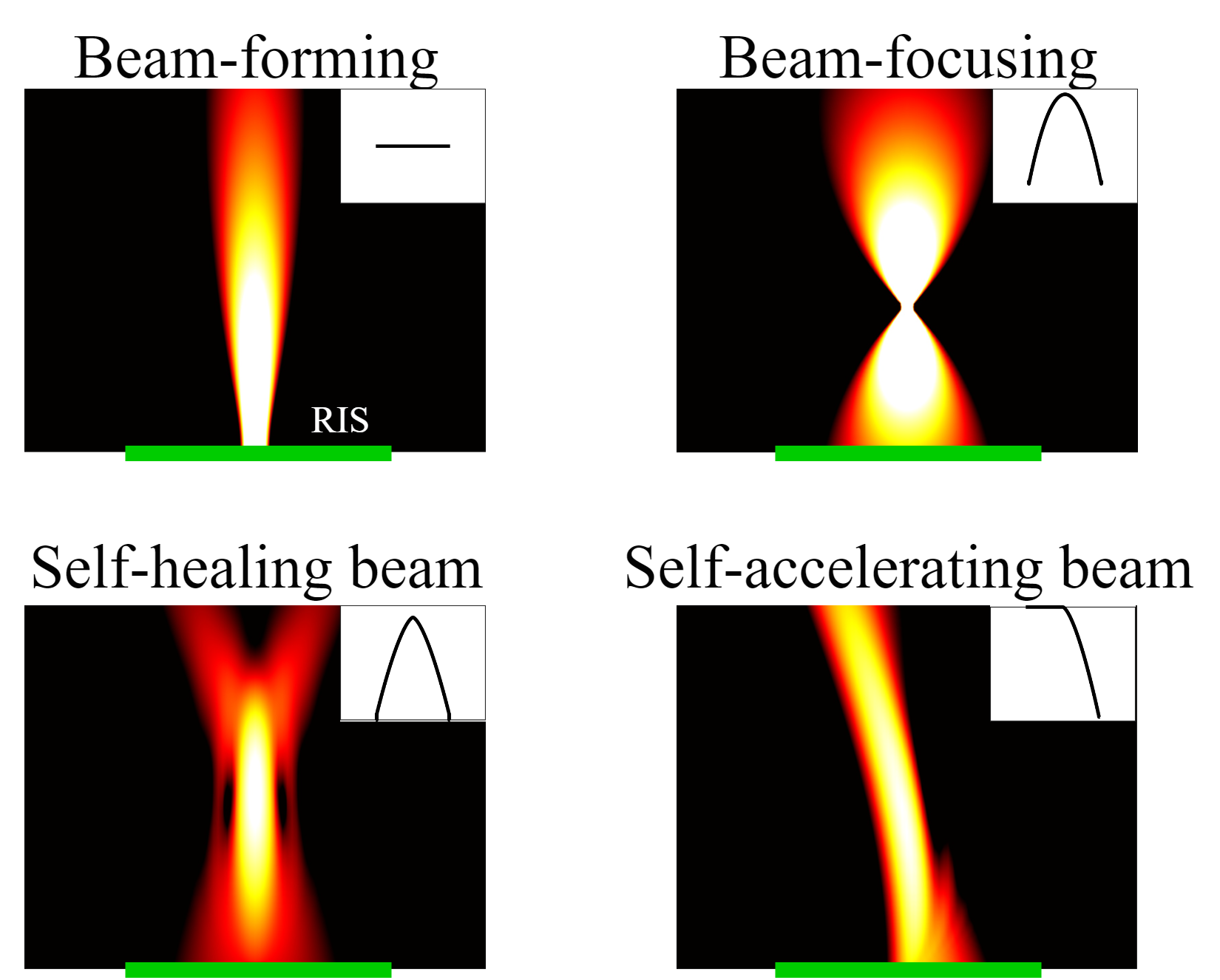}
\caption{Beam-forming and other advanced beam profiles. The insets show cross-section of the phase profile of the RIS elements that generates each beam. The operation frequency is 150 GHz ($\lambda=2$ mm) and the RIS (marked with the green line) consists of 1000 $\times$ 1000 elements of size $\lambda/2$.}
\label{fig:fig1}
\end{figure}
%
%
%
\noindent It is well known that the waves radiated by antennas evolve into the far-field roughly at the Fraunhofer distance, $2D^2/\lambda$ which is determined by the size of the antenna $D$ relative to the wavelength $\lambda$. For conventional antennas, e.g. phased arrays, the size of the radiating element (or radiating aperture) extends from a fraction of a wavelength up to only a few wavelengths and, therefore, the radiated waves quickly enter the far-field. With RISs, though, this intuitive picture can change. Owing to their large surface, usually extending up to several wavelengths, waves reflected by the RIS may evolve under a new aperture, i.e. a new Fraunhofer distance. For example, a wave propagating in the far-field of the transmitter and impinges on a RIS, illuminating the entire RIS surface, is steered under a new aperture determined by the size of the RIS (this is typically referred to as \textit{full illumination}). On the other hand, a pencil beam that partially illuminates a large RIS area is steered with a new aperture determined by the size of the beam footprint on the RIS (this is typically referred to as \textit{partial illumination}). As a result, the transition from the near- to the far-field can be much farther than what is intuitively expected at a certain frequency and can be controlled by the size of the RIS relative to the footprint of the incident beam \cite{Stratidakis2021a}. This flexibility to adjust the Fraunhofer distance can enable new functionalities, such as focusing and generally advanced beams that require relatively large apertures with tailored phase response. \\
The RIS transforms the incident beam to a reflected beam with desired characteristics. The profile of the reflected beam is controlled by the phase, $\phi_{RIS}$, imposed by the RIS elements on the incoming wavefront, i.e. the incident field, $E_i$, is transformed to the reflected field as $E_r = E_i\exp(j \phi_{RIS})$. Depending on the functional form of $\phi_{RIS}$, different types of beams can be generated by the RIS \cite{Droulias2022}. In Fig. \ref{fig:fig1} four beam profiles are presented, corresponding to \textbf{beamforming}, \textbf{beam-focusing}, \textbf{self-healing}, and \textbf{self-accelerating} operations. These profiles can be advantageous to a number of applications, including but not limited to beam-tracking, blockage avoidance and power transfer. \\
\indent \textbf{Beamforming} is the most fundamental of the functionalities of RISs \cite{Björnson2020} and is typically used to steer the incident beam towards the desired direction. This is achieved by imposing linear phase shifts on the RIS elements, i.e. $\phi_{RIS}= k(sin\theta_i-sin\theta_r)x$, where $k$ is the free-space wavenumber, $\theta_i$, and $\theta_r$ denote the angles of incidence and reflection on the steering plane, respectively, and $x$ is the Cartesian coordinate; in this example the RIS is on the $xy$ plane, steering waves on the $xz$ plane.
Beamforming with a RIS can be used to increase the received power at the direction of the user, establish an additional LoS link, or track the user.
The size of the RIS affects the reflected beam in two ways. 
First, the beam-width of the reflected beam depends on the size of the footprint of the incident beam on the RIS or the size of the RIS, whichever is smaller, which in turn depends on the beam-width of the incident beam. As a result, as the footprint of the incident beam increases under partial illumination conditions, the beam-width of the reflected beam changes until, passing onto full illumination conditions, the footprint surpasses the RIS size and the reflected beam converges to a shape determined by the size and the shape of the RIS \cite{Stratidakis2021a}.
Second, the size of the RIS/footprint also affects the Fraunhofer distance. For example, a large footprint increases the Fraunhofer distance and can potentially ``move" the user equipment (UE) in the near-field. On the other hand, a small footprint decreases the Fraunhofer distance and can ``move" the UE to the far-field. The Fraunhofer distance can be effectively controlled through the gain of the transmitter, which tunes the size of the beam-footprint on the RIS. Controlling the Fraunhofer distance is fundamental as the optimization of beamforming in terms of received power depends on it. As shown in \cite{Stratidakis2021a}, maximizing the received power at a certain distance requires a specific RIS size or footprint that sets the Fraunhofer distance at the RIS-UE distance, i.e. that effectively places the UE at the near-to-far-field transition of the RIS. Moreover, under partial illumination, the beam in its near-field undergoes negligible diffraction, essentially propagating with constant peak power, which is ideal for power transfer applications. \cite{Tran2021} \\ 
\indent \textbf{Beam-focusing} is the ability of a RIS to focus the power of the reflected beam to a small area, similarly to how a lens focuses light onto a focal point.
This can be achieved simply by imposing a parabolic phase profile, $\phi_{RIS}$, on the RIS elements, i.e. of the form $-kr^2/2f_0$, where $r=\sqrt{x^2+y^2}$ is the radial coordinate in the RIS $(x,y)$ plane, and $f_0$ is the focal distance.
Beam-focusing can be used to estimate the location of the UE, to maximize the received power, as well as for wireless power transfer applications. The power is maximized at the focal point, and the locus of the spatial locations with half-maximum (or 3dB) power marks an ellipsoid centered at the focal point and oriented with its major axis along the beam propagation direction. Its size depends on the incident beam footprint, as well as the RIS-focal point distance (focal distance). In principle, the larger the footprint on the RIS, the smaller the size of the 3dB ellipsoid and, consequently, the higher the resolution in localization-related applications and the higher the maximum power at the focal point. \\
\indent \textbf{Self-healing beams} are a type of diffraction-free beams with the ability to reconstruct themselves when faced with an obstacle \cite{Aiello2017}. This is possible, because rays from the edge of the beam replace the blocked rays as the beam propagates. Therefore, self-healing beams are ideal for counteracting the effect of blockage. Such beams require a power-law phase profile of the form $-kCr^\gamma$, where $\gamma$ and $C$ are design parameters (constants) \cite{Efremidis2019}. Theoretically, although such beams require an infinitely large array to be generated properly \cite{Inserra2022}, they can be efficiently generated with finite apertures, as shown in Fig. \ref{fig:fig1}. The effectiveness, as well as the distance of beam reconstruction depend on the size of the obstacle relative to the RIS aperture. In principle, the larger the RIS, the faster and better the beam recovers. \\
\indent \textbf{Self-accelerating beams} are also diffraction-free, but unlike self-healing beams, they bend with the propagation distance. They require a power-law phase profile of the form $-kCx^\gamma$, where $\gamma$ and $C$ are constants that control the bending radius. Such beams also possess the ability to reconstruct after an obstacle like all diffraction-free beams \cite{Efremidis2019,Aiello2017} and, therefore, such beams are also ideal for counteracting the effect of blockage. In principle, the larger the RIS, the more the beam can be engineered to bend. With knowledge of the wireless environment, the location of the UE, and the beam propagation, self-accelerating beams can establish connection between points otherwise inaccessible with conventional beamforming.
\begin{table}
\begin{tabular}{||c | c | c | c | c||} 
 \hline
 Properties & \begin{tabular}{@{}c@{}} Beam-\\steering \end{tabular}  & \begin{tabular}{@{}c@{}} Beam-\\focusing \end{tabular} & \begin{tabular}{@{}c@{}} Self-\\healing \end{tabular} & \begin{tabular}{@{}c@{}} Self-\\accelerating \end{tabular} \\ [0.5ex] 
 \hline\hline
 Perceptive & \checkmark & \checkmark & - & - \\ 
 \hline
 \begin{tabular}{@{}c@{}} Blockage\\resilience \end{tabular} & - & - & \checkmark & \checkmark \\
  \hline
 \begin{tabular}{@{}c@{}} Energy\\efficiency \end{tabular} & \checkmark & \checkmark & - & - \\
 \hline
\end{tabular}
\caption{Capabilities of the four beam profiles.}
\label{tab:tab1}
\end{table}
%
%
\section{Advanced wavefront engineering techniques}
%
\begin{figure}[t!]
\centering
\includegraphics[width=\linewidth]{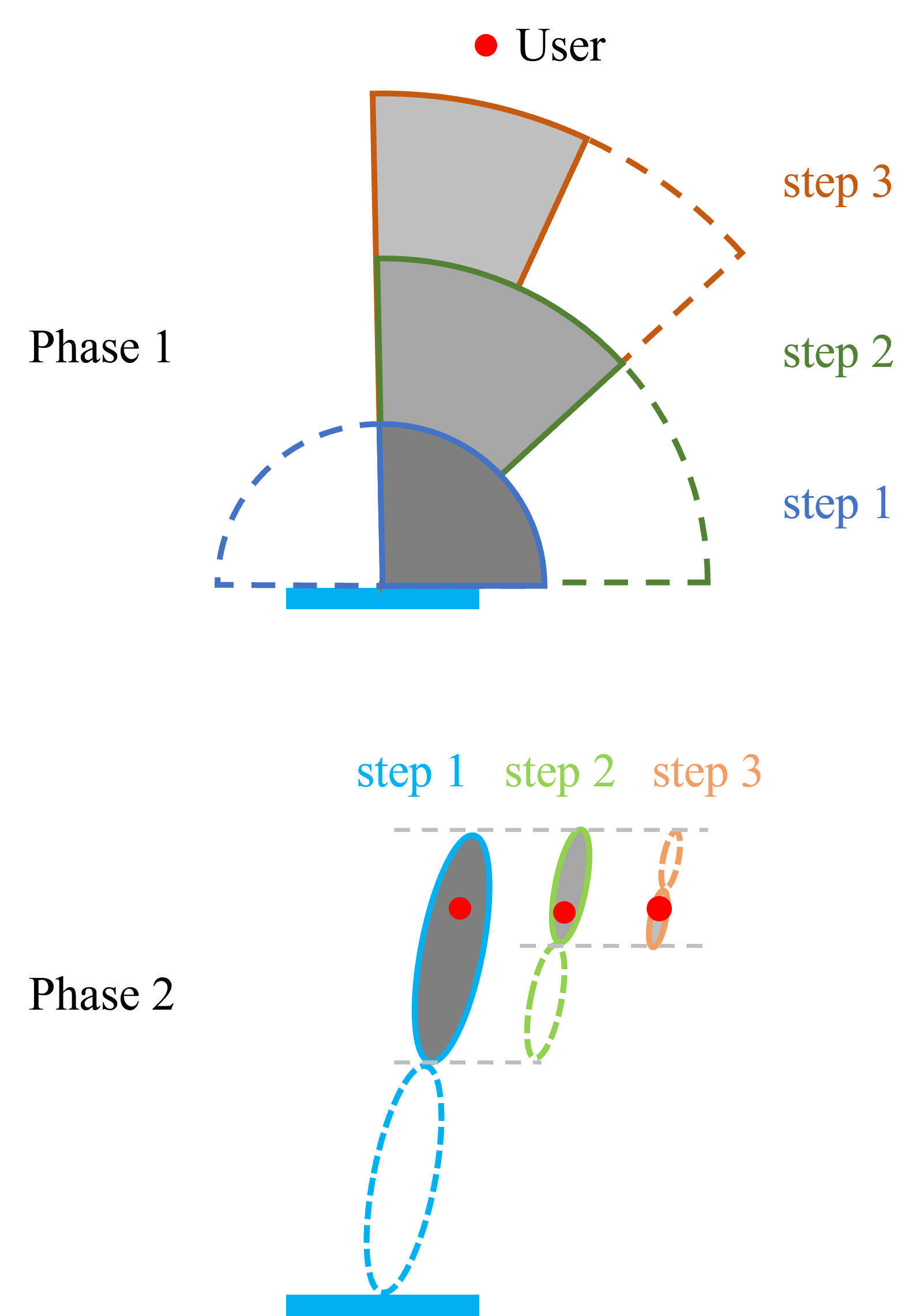}
\caption{Hierarchical RIS-assisted localization. The shaded regions mark the areas where the user location is successfully identified at each hierarchical level.}
\label{fig:fig2}
\end{figure}
%
\subsection{Perceptive}
\noindent Wireless communications have been so far operating in the far-field of the radiating elements. With the incorporation of RISs, though, the near-to-far-field transition can be engineered via the footprint size and, therefore, new techniques that can take advantage of this ability can be implemented. These techniques are designed to enable the three main properties that future networks are expected to have, namely being perceptive, resilient and efficient. Table \ref{tab:tab1}, summarizes these properties, and the relevant beam profiles to realize them.
Wireless communication environments are highly volatile. This is especially true in dense networks with highly directional antennas operating at high frequency bands, as future networks are expected to be. As a result, such networks must be perceptive within the wireless environment, to be able to adjust to all scenarios.

\textit{\underline{Hierarchical RIS-assisted localization}}: 
The ability of the RIS to focus the power of the reflected beam to a small area is ideal for localization. Due to the small size of the focal area, searching the entire area of interest with beam-focusing alone may require prohibitively large overhead and, therefore, searching can be assisted by hierarchical beam-tracking. The whole procedure can be split in two phases as shown in Fig. \ref{fig:fig2}.

In phase 1, beam-tracking is performed with beam-forming through the RIS, in order to estimate the direction of the UE with respect to the RIS. In phase 2, ranging is performed with beam-focusing through the RIS, in order to estimate the distance of the UE from the RIS, along the direction identified in phase 1. With information about both the direction and the distance, the location of the UE is fully determined. Both phases follow are hierarchical and binary tree structured.
Essentially, the area of interest (e.g. the semi-infinite space in front of the RIS) is represented by a polar grid, with the RIS at its origin. In phase 1 the angular sector where the user is located is identified and in phase 2 the radial distance of the UE is estimated within the available resolution of the focal spot.

\underline{Phase 1: Beam-forming (beam-tracking)}: Conventional beam-tracking is performed in the far-field, where the beam spreads as the propagation distance increases \cite{Stratidakis2022b}. In general, the access point (AP) scans a number of predefined directions which depends on the number of antenna elements. The number of directions to scan can be reduced significantly using a hierarchical codebook that follows the binary tree structure \cite{Xiao2016}. The hierarchical codebook is comprised of $log_2N$ levels, where $N$ is the number of antenna elements. In the first level, the AP divides the area to two sectors and finds the UE in one of them. In each subsequent level, the AP divides the previous beam, where the UE was found, to two more sectors. This continues until the highest level, where the narrowest beam is reached. With this method, the number of directions to scan with exhaustive search changes from $N$ to $2 log_2(N)$, and this difference becomes more significant with increasing $N$. Because larger antenna aperture pushes the Fraunhofer distance farther, the choice for the maximum $N$ defines also the minimum distance from the antenna where this technique can be successfully applied.
The same beam-tracking method can be used with a RIS. Unlike antenna arrays where the antenna elements are directly fed by currents, the RIS elements are excited by the incident wave. By controlling the properties of the incident wave, the excitation conditions of the RIS elements can be tuned and, therefore, the adjustment of the number of RIS elements is mainly performed by controlling the AP antenna gain and therefore, the beam-footprint on the RIS.

\underline{Phase 2: Beam-focusing (ranging)}: With beam-focusing, most of the power is concentrated along the direction of the reflected beam.  
With knowledge from phase 1 of the angular sector where the UE is located, the elliptical shape and controllable size of the focal area make beam-focusing ideal for hierarchical localization. This approach can be implemented as follows. 
The RIS divides the angular sector from phase 1 in two distance-based sectors with beam-focusing at two large focal areas that cover more than half of the maximum distance each. The maximum distance is chosen at will and the focal areas can be chosen to overlap to avoid blind spots. After the user is found in one of those areas, the RIS divides that area in two smaller ones by adjusting the footprint size and focal distance. This process is repeated until the desired accuracy of the localization is achieved, as shown in Fig. \ref{fig:fig2}. \\
%
\begin{figure}[t!]
\centering
\includegraphics[width=\linewidth]{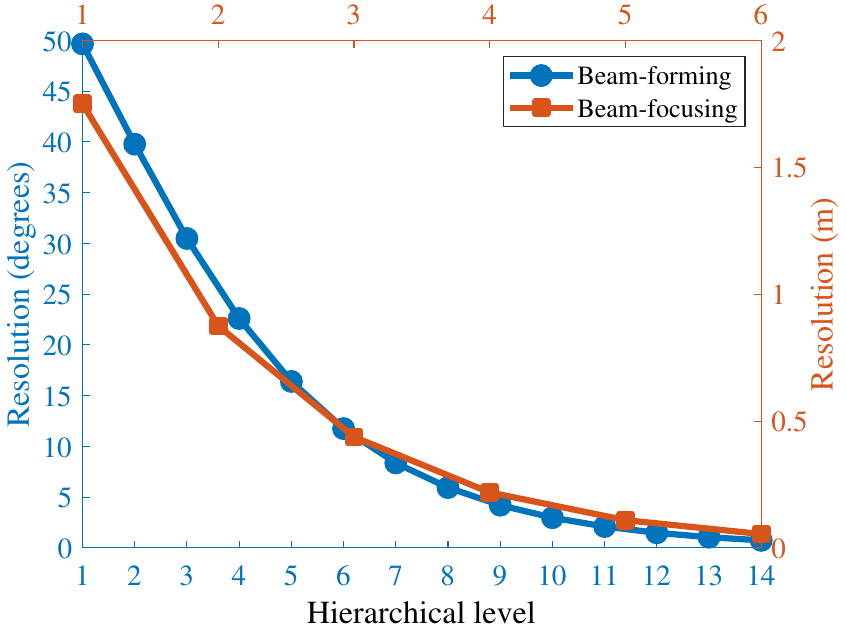}
\caption{Resolution vs hierarchical level for the two phases of the hierarchical RIS-assisted localization scheme. For beam-focusing in Phase 2, the maximum distance is $5$ m.}
\label{fig:fig3}
\end{figure}
%
%
The resolution for phase 1 depends on the 3dB (angular) width of the beam, and for phase 2 on the size, and more specifically, the major radius (since it has elliptical shape) of the focal area. In Fig. \ref{fig:fig3}, the resolution of each phase is presented as a function of its hierarchical level, for localizing a UE that is within a maximum radius of $5$ m from the RIS. For both phases, a higher hierarchical level results in refined resolution. For phase 1 the maximum number of hierarchical levels depends on the maximum desired angular resolution. For phase 2 the maximum number of hierarchical levels depends on the maximum desired radial resolution, as well as on the maximum distance that is to be scanned. For example, if the maximum distance is doubled to be $10$ m, one more level is added with half the resolution of the previous level, where the maximum distance was $5$ m, and so on.\\
\indent The highest hierarchical level defines the maximum number of areas (direction-wise or distance-wise) that are scanned to find the user with the desired resolution. As the approach is binary-tree structured, with each hierarchical level two areas are scanned. Therefore, Fig. \ref{fig:fig3} implies that there is a trade-off between how fast the UE location is to be determined and with what accuracy: finer resolution requires higher hierarchical levels, i.e. additional steps in the binary search procedure.
The search can be accelerated with prediction. In the abscence of prediction all hierarchical levels are necessarily utilized to find the user. However, with enough samples about the location of the UE, the AP can start predicting the next location of the user in relation to the location of the RIS and scan a small area around the predicted location using both phases but with fewer number of hierarchical levels. Prediction of the user's location in the next timeslot enables the AP to start the search from a higher hierarchical level in both phases, resulting in reduced overhead and required time to find the UE location. \\
\indent The size of the focal area and, hence, the highest achievable resolution in phase 2, depends on the size of the beam-footprint on the RIS. Overall, the larger the footprint on the RIS, the highest the achievable resolution. Fig. \ref{fig:fig4} shows the required footprint radius to achieve a fixed focal area size of major radius $0.1, 0.2, 0.5$ and $1$ m at focal distance from $2$ to $10$ m. As the focal distance increases, the footprint radius of the beam on the RIS must increase as well to ensure that the size of the focal area stays the same. Otherwise, the focal area is elongated along the direction of the reflected beam.
\begin{figure}[t!]
\centering
\includegraphics[width=\linewidth]{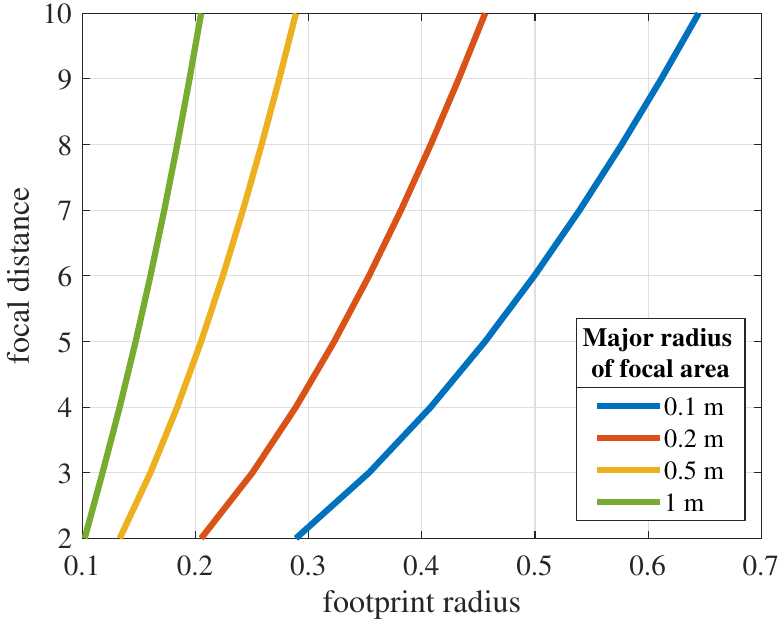}
\caption{Focal distance vs RIS footprint size for achieving the focal area size shown in the inset.}
\label{fig:fig4}
\end{figure}
\subsection{Resilient}
\noindent In recent years, more and more real time applications emerge, such as video conferences and live streaming. As a result, resilience to outage is becoming increasingly more significant, calling for networks that can avoid outages and quickly restore the link.

In high frequencies, one significant cause of outage is blockage. For example, at $73.5$ GHz the power losses can reach $40$ dB \cite{MacCartney2016}. At higher frequencies the losses are higher. To this end, the use of RISs has been proposed, in order to offer an additional LoS link and avoid obstacles entirely with low power requirements and without extra AWGN. However, in environments with many obstacles (humans, furniture, etc.) adding more RISs does not necessarily increase the LoS probability. Therefore, new ways to resist blockage are required. 

\subsubsection{\underline{Resilience against blockage}}
\begin{figure}
\centering
\includegraphics[width=\linewidth]{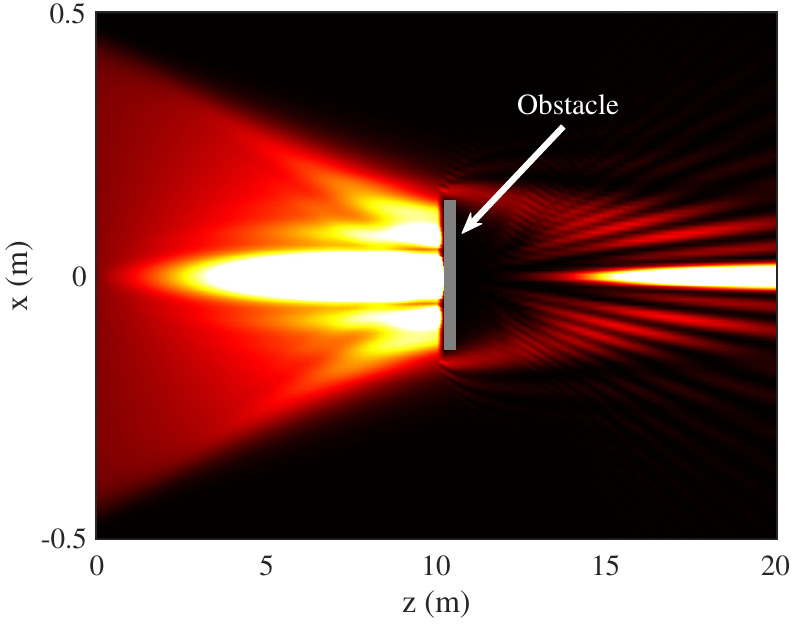}
\caption{A self-healing beam reconstructs after encountering an obstacle.}
\label{fig:fig5}
\end{figure}

Non-diffracting beams, such as \textbf{self-healing} and \textbf{self-accelerating} beams can reconstruct behind obstacles, making them resilient to blocked links, as long as the size of the obstacle is smaller than the beam and a part of the energy is not blocked as shown in Fig. \ref{fig:fig5}. Such beams are ideal candidates for resilience against blockage.

%
\subsubsection{\underline{Blockage avoidance}}
\begin{figure}
\centering
\includegraphics[width=\linewidth]{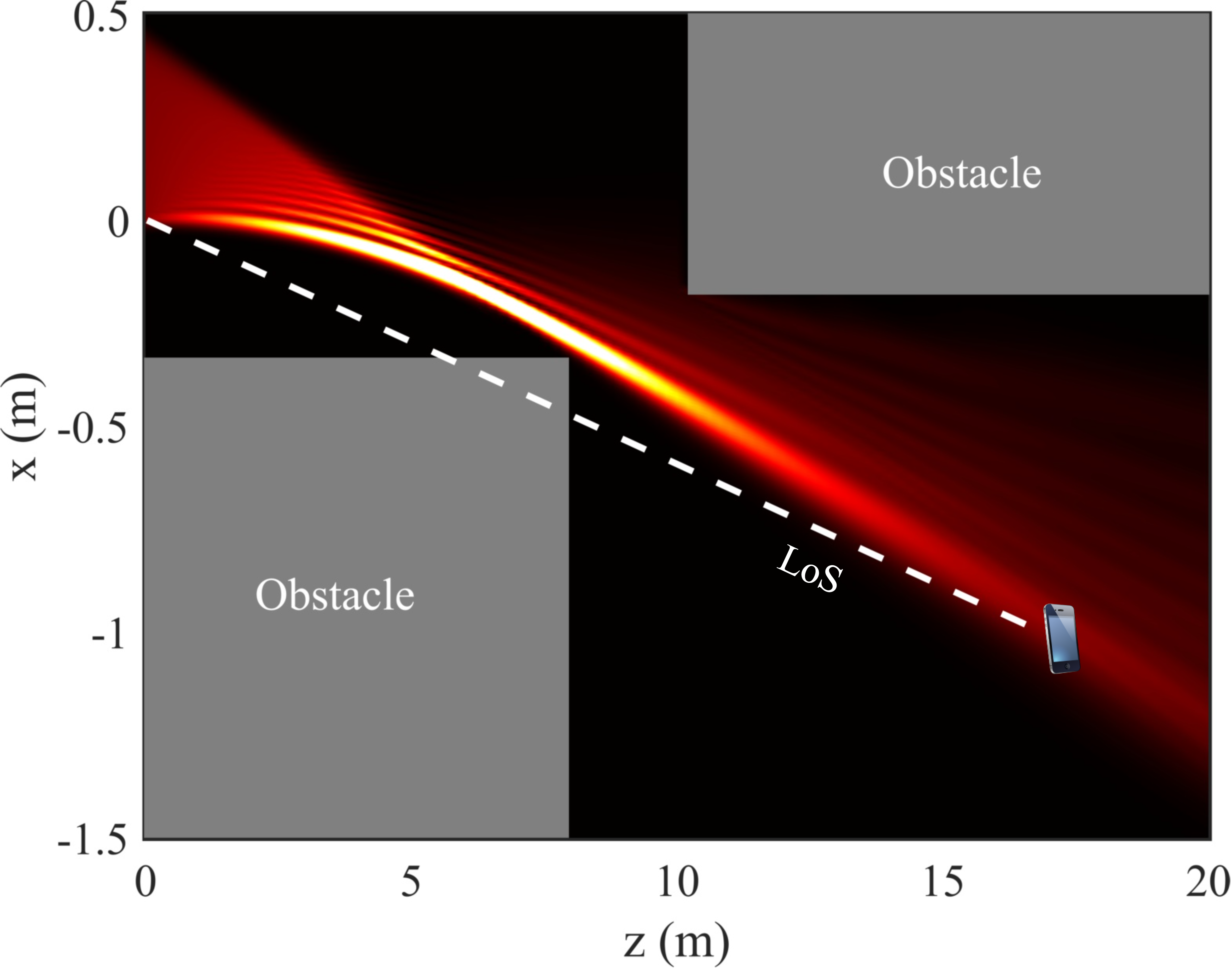}
\caption{A self-accelerating beam avoids the obstacles, reaching the UE beyond the LoS.}
\label{fig:fig6}
\end{figure}

\textbf{Self-accelerating beams} have the ability to ``bend" with the propagation distance, and possibly restore the LoS when the LoS of the AP-UE and AP-RIS-UE links are blocked. This ability offers a way to avoid obstacles without additional APs, relays or RISs. It can be especially useful in cases with static obstacles in indoor scenarios or obstacles that unexpectedly interrupt the LoS of the RIS-UE link in outdoor scenarios. This possibility is demonstrated in Fig. \ref{fig:fig6}, where a self-accelerating beam is used to avoid two large obstacles and reach a UE at a position beyond the RIS-UE LoS, i.e. not accessible with conventional beamforming. 

\subsection{Efficient}
\noindent The main reasoning behind the expected low energy consumption of dense networks is that as each AP covers a small area, 
less power will be overall required. However, depending on the density of the networks, the energy consumption can still be quite high. For that reason, the energy efficiency must be further improved. This can be achieved directly by reducing the energy consumption of the networks, at the expense of reduced transmitted power and, hence, reduced signal-to-noise ratio (SNR). To maintain high SNRs with reduced power consumption, new functionalities, such as power transfer or sensing (e.g. localization), can be incorporated with communications.

\subsubsection{\underline{Energy efficient communications}} 
RISs are nearly passive surfaces and, therefore, give promise for energy efficient communications, as opposed to power-hungry relays or the deployment of additional APs. In view of their potential to be exploited for advanced wavefront engineering, their efficiency can be further boosted with different beam profiles serving different scenarios. For example, beam-forming with high gain concentrates the power to a desired direction. Self-healing and self-accelerating beams suppress or even eliminate the effect of blockage reducing the need for different routes, more hops and more APs, RISs or relays, all of which increase the energy consumption. 
Beam-focusing concentrates the power of the reflected beam at a small area with controllable size. The size of the focal area depends on the beam-footprint size on the RIS, with two important implications: (a) the power at the focal point can be significantly increased by enlarging the incident beam's footprint on the RIS, while keeping the power of the incident beam constant; (b) the same power at the focal point can be achieved with beams of lower power and larger footprint on the RIS. Therefore, focused beams are ideal for energy efficient applications, as they can either amplify the received power with no additional power needed or they can offer the same received power with significantly reduced power consumption. Importantly, due to the relatively small size of the focal area, the interference with other focal areas is practically negligible. Therefore, beam-focusing is ideal for high signal-to-interference ratio at the receiver with low transmitted power.

\subsubsection{\underline{Power transfer}}
Wireless power transfer is an application that is rapidly gaining ground in recent years. Although it can be performed with all beam profiles, some are better suited for this application than the others. \textbf{Beam-forming} with partially illuminated RIS offers uniform beams from the near-to the far-field without near-field oscillations (as would occur with fully illuminated RIS). Importantly, at distances shorter than the Fraunhofer distance the beam hardly diffracts. Hence, such beams can provide almost constant power along their propagation direction for areas of interest within such distances \cite{Feng2022}. Last, the ability of \textbf{beam-focusing} to concentrate the reflected beam power at a desired area with adjustable size is ideal for power transfer applications.

\section{Conclusions}
\noindent To this day, communication networks are designed to operate in the far-field. The need for agility, programmability and recofigurability pushes the operating frequency to higher bands and imposes the need for large antenna arrays and RISs, consequently bringing communications in the near-field. The large area of the RIS and the multitude of scatterers it offers, renders the RIS ideal for advanced wavefront engineering, where the incident beam is transformed into a non-trivial reflected beam that is able to address the challenges of high frequencies more efficiently than conventional beam-forming.
In this paper, the potential of advanced wavefront engineering was investigated for RIS-assisted networks, with the purpose of making the networks perceptive to changes in the wireless communication environment, resilient to outages, and energy efficient. 
Three advanced wavefront engineering beam profiles are demonstrated along with RIS-assisted beam-forming, the beam-focusing that concentrates the power at a small area, the self-healing beam profile that is resilient to obstacles, and the self-accelerating beam profile that can avoid obstacles.
The design specifications to reach their full potential are discussed, and some techniques are demonstrated in an effort to make the wireless communications networks perceptive, resilient and efficient.

\section*{Acknowledgment}
\noindent This work was supported by the European Commission’s Horizon Europe Programme under the Smart Networks and Services Joint Undertaking TERA6G project (Grant Agreement 101096949) and INSTINCT project (Grant Agreement 101139161).

\bibliographystyle{IEEEtran}
\bibliography{IEEEabrv,References}

\end{document}